\documentclass[aps,pra,superscriptaddress,reprint,floatfix,longbibliography,showpacs,amsfonts,amsmath,amssymb]{revtex4-1}
\usepackage{graphicx}
\usepackage{hyperref}
\usepackage{xr-hyper}
\usepackage{upgreek}
\usepackage{epstopdf}
\usepackage[note-name=, use-sort-key = false]{notes2bib}
\usepackage{color}
\usepackage{comment}

\begin{document}
\title{In-plane and out-of-plane optical response of the nodal-line semimetals ZrGeS and ZrGeSe}

\author{J. Ebad-Allah}
\affiliation{Experimentalphysik II, Augsburg University, 86159 Augsburg, Germany}
\affiliation{Department of Physics, Tanta University, 31527 Tanta, Egypt}
\author{S. Rojewski}
\affiliation{Experimentalphysik II, Augsburg University, 86159 Augsburg, Germany}
\author{Y. L. Zhu}
\affiliation{2D Crystal Consortium, Materials Research Institute, Pennsylvania State University, University Park, PA 16802, USA}
\affiliation{Department of Physics, Pennsylvania State University, University Park, Pennsylvania 16802, USA}
\author{Z. Q. Mao}
\affiliation{2D Crystal Consortium, Materials Research Institute, Pennsylvania State University, University Park, PA 16802, USA}
\affiliation{Department of Physics, Pennsylvania State University, University Park, Pennsylvania 16802, USA}
\affiliation{Department of Materials Science and Engineering, Pennsylvania State University, University Park, Pennsylvania 16802, USA}
\author{C. A. Kuntscher}
\email{christine.kuntscher@physik.uni-augsburg.de}
\affiliation{Experimentalphysik II, Augsburg University, 86159 Augsburg, Germany}

\begin{abstract}
Polarization-dependent reflectivity measurements were carried out over a broad frequency range on single-crystalline ZrGeSe and ZrGeS compounds, which are closely related to the prototype nodal-line semimetal ZrSiS. These measurements revealed the strongly anisotropic character of both ZrGeSe and ZrGeS, with a reduced plasma frequency for the out-of-plane direction {\bf E}$\| c$ as compared to the in-plane direction {\bf E}$\| ab$.
For {\bf E}$\| ab$ the optical conductivity spectrum consists of two Drude terms followed by a shoulder or plateau-like behavior and a distinct U shape at higher energies, while for {\bf E}$\| c$ one Drude term is followed by a peak-like behavior and the U shape of the profile is less developed.
Under external pressure, two prominent excitations appear in the out-of-plane optical conductivity spectrum of ZrGeSe, whose frequency position and oscillator strength show a weak anomaly at $\sim$3~GPa. Overall, the pressure-induced changes in the profile of the {\bf E}$\| c$ conductivity spectrum are much enhanced above $\sim$3~GPa.
We compare our results to those recently reported for ZrSiS in a quantitative manner.

\end{abstract}

\maketitle

\section{Introduction}

The ZrSiS-type family of square-net materials with the general formula {\it WXY} (with $W$ = Zr, Hf; $X$ = Si, Ge, Sn; $Y$ = O, S, Se, Te)
has been extensively investigated experimentally and theoretically in recent years, since they provide an ideal platform for
exploring exotic states of quantum matter, namely the nodal-line semimetal phase \cite{Neupane.2016,Schoop.2016,Topp.2016,Hu.2016,Klemenz.2018,Kirby.2022}
and novel two-dimensional Dirac states protected by non-symmorphic symmetry \cite{Young.2015}.

The {\it WXY} compounds crystallize in the PbFCl-type structure in the tetragonal space group $P$4/$nmm$.
Their crystal structure consists of slabs with five square nets, with a stacking sequence [$Y$-$W$-$X$-$W$-$Y$]  \cite{Wang.1995,Sankar.2017}. The relatively weak bonding between two adjacent slabs is of van der Waals-type. The lattice parameters of the studied materials ZrSiS, ZrGeSe, and ZrGeS are listed in Table~\ref{tab:optical parameters}.
ZrSiS, which is considered as the prototype nodal-line semimetal, shows exceptional physical properties, like extremely large magnetoresistance and high carrier mobility \cite{Sankar.2017,Lv.2016,Singha.2017}.
Its electronic band structure contains
lines of Dirac nodes at the Fermi energy $E_F$, which form a cage-like structure around the $\Gamma$-point and result in a diamond-shaped Fermi surface \cite{Schoop.2016}. The energy range of the linear dispersion of Dirac bands in ZrSiS is exceptionally large as compared to other known Dirac materials, namely 2~eV above and below the Fermi energy $E_F$ in some regions of the Brillouin zone.
Besides the Dirac nodes close to $E_F$, there are Dirac-like band crossings at energies 0.5 - 0.7 eV below $E_F$ at the $X$ and $R$ point of the Brillouin zone (BZ), which are 2D in character and protected by the non-symmorphic symmetry against gapping \cite{Schoop.2016}. A recent ARPES study suggested that these Dirac nodes lie on a nodal plane \cite{Fu.2019}.
By the choice of $X$ and $Y$ element in {\it WXY}, the dimensionality of the compound can be tuned between 3D and quasi-2D: Starting from the most-studied ZrSiS, the slab thickness increases significantly along the $c$ axis for ZrSiSe and even more for ZrSiTe, due the increased ionic radius of the chalcogene element $Y$. Therefore, the interlayer bonding is decreased in ZrSiTe \cite{Wang.1995}, leading to a reduction of the structural dimensionality to be close to 2D.
A similar tuning of the structural dimensionality can be achieved by substitution of the $X$ element, but not as straightforward like for the $Y$ substitution, as in a previous optical study  \cite{Ebad-Allah.2019}. The structural dimensionality is monitored via the $c$/$a$ ratio of the lattice parameters, where $c$ is the out-of-plane and $a$ is the in-plane lattice parameter \cite{Wang.1995} and is expected to severely affect the electronic band structure in terms of energetic shifts of bands \cite{Topp.2016}.

Previous frequency-dependent reflectivity measurements on ZrSiS for the polarization of the incident radiation along ({\bf E}$\| ab$) and perpendicular ({\bf E}$\| c$) to the layers revealed a highly anisotropic optical response, in very good agreement with density-functional-theory (DFT) calculations \cite{Ebad-Allah.2021}. Corresponding optical measurements under external pressure showed only small pressure-induced changes in the in-plane optical conductivity, whereas strong effects were observed in the out-of-plane optical conductivity spectrum of ZrSiS, with the appearance of two prominent excitations. Since a pressure-induced structural phase transition could be ruled out based on x-ray diffraction data, the pressure-induced changes were suggested to be of electronic origin. Moreover, DFT calculations showed that the pronounced peaks in the {\bf E}$\| c$  optical conductivity spectra of ZrSiS cannot be attributed to electronic correlation and electron-hole pairing effects  \cite{Ebad-Allah.2021}.

In this paper, we study the polarization-dependent optical response of two other members of the compound family {\it WXY}, namely ZrGeSe and ZrGeS. Similar to ZrSiS, we find a highly anisotropic character of the optical conductivity with a much reduced metallic character along the out-of-plane direction. In particular, the pressure-induced appearance of two new excitations in the out-of-plane optical response is confirmed in the compound ZrGeSe. We compare the optical conductivity of the various compounds under ambient and high-pressure conditions in a quantitative manner.

\begin{table}[t]
    \centering
   \begin{tabular}{c|ccc|ccc}
    \hline
      & ZrSiS & ZrGeSe & ZrGeS &  &  &  \\
     \hline
     $a$         & 3.544  & 3.706 &  3.626 &    &     &   \\
     \hline
     $c$         &  8.055 & 8.271 &  8.019 &    &    &   \\
     \hline
      $c$/$a$    & 2.273  & 2.232  & 2.212  &    &    &    \\
       \hline

      & &  {\bf E}$\| ab$  & & & {\bf E}$\| c$  & \\
      & ZrSiS & ZrGeSe & ZrGeS & ZrSiS & ZrGeSe & ZrGeS \\
     \hline
      $\omega_{pl}^{scr}$ (eV)     & 1.07  & 0.99  & 1.07  & 0.47  & 0.44  & 0.36  \\
      \hline
    $\omega_{pl}$ (eV)   & 3.17 & 2.96 & 3.42 & 1.08 & 1.06 & 0.95 \\
    \hline
    L4 position (eV)    & 1.44  & 1.30  & 1.39  & ---  & ---  & ---  \\
   \hline
   1.5 GPa  &   &   &   &   &   & \\
    F1 position (eV)   & ---  &  --- & ---  & 0.93  & 0.82  &   \\
    F2 position (eV)   & ---  &  --- & ---  & 1.70  & 1.52  &   \\
   \hline
   6.0 GPa   &   &   &   &   &   & \\
    F1 position (eV)   & ---  &  --- & ---  & 0.80  & 0.61  &   \\
    F2 position (eV)   & ---  &  --- & ---  & 1.70  & 1.58  &   \\
   \hline
\end{tabular}
 \caption{Lattice parameters $a$ and $c$, and the ratio $c$/$a$ from Ref.\ \cite{Haneveld.1957}, and optical parameters of ZrGeSe, ZrGeS, and ZrSiS. Results for ZrSiS have been reported in Ref.\ \cite{Ebad-Allah.2021}.}
 \label{tab:optical parameters}
\end{table}

\section{SAMPLE PREPARATION AND EXPERIMENTAL DETAILS}

ZrGeSe and ZrGeS single crystals were grown by chemical vapor transport method \cite{Hu.2017a,Hu.2017} and characterized by x-ray diffraction and energy-dispersive x-ray spectroscopy, in order to ensure phase-purity and crystal quality.

An infrared microscope (Bruker Hyperion), equipped with a 15$\times$ Cassegrain objective, coupled to a Bruker Vertex 80v FT-IR spectrometer was used for carrying out the polarization-dependent reflectivity measurements at ambient and high pressure, both at room temperature.
The ambient-pressure reflectivity measurements were carried out in the frequency range 200-23.000 cm$^{-1}$ on surfaces cleaned with alcohol. As reference, a commercial Al mirror was used. The reflectivity results for ZrSiS are consistent with those in Ref.\ \cite{Schilling.2017}, and therefore the assumption of normal incidence of the electromagnetic radiation from the Cassegrain objective to the sample is valid.
The optical functions were obtained by Kramers Kronig (KK) transformation of the reflectivity spectra. Hereby, the reflectivity data were extrapolated to low frequencies based on a Drude-Lorentz fit, while for the high-frequency extrapolation we used the x-ray atomic scattering functions \cite{Tanner2015}.

For the generation of pressures up to 6.0~GPa a Diacell CryoDAC-Mega (AlmaxEasyLab) diamond-anvil cell (DAC) was used. The out-of-plane surface of ZrGeSe single crystals, with the size of about 210$\times$190$\times$80 $\upmu$m$^{3}$, was carefully cleaned with isopropanol and loaded in the hole of a CuBe gasket inside the DAC. Finely ground CsI powder served as quasi-hydrostatic pressure transmitting medium and to ensure the sample-diamond interface during the optical measurements. The pressure was determined {\it in situ} using the ruby luminescence technique \cite{Mao.1986,Syassen.2008}.
The pressure-dependent reflectivity spectra $R_\text{s-d}$ at the sample-diamond interface for frequencies 300-9.000~cm$^{-1}$ were obtained according to $R_\text{s-d}(\omega)$=$R_\text{gasket-dia}(\omega)$$\times$$(I_\text{s}(\omega)/I_\text{gasket}(\omega))$,
where $I_\text{s}(\omega)$ is the intensity of the radiation reflected from the sample-diamond interface, $I_\text{gasket}(\omega)$ the intensity reflected from the CuBe gasket-diamond interface, and $R_\text{gasket-dia}(\omega)$ the reflectivity of the gasket material for the diamond interface. The $R_\text{s-d}$ spectra in the frequency range 9000-19.000 cm$^{-1}$ were calculated according to
$R_\text{s-d}(\omega)$=$R_\text{dia}$$\times$$(I_\text{s}(\omega)/I_\text{dia}(\omega))$,
where $I_\text{s}(\omega)$ is the intensity reflected at the interface between the sample and the diamond anvil, $I_\text{dia}(\omega))$ is the intensity reflected from the inner diamond-air interface of the empty DAC, and $R_\text{dia}$=0.167 is the reflectivity of diamond, which was assumed to be pressure independent \cite{Eremets.1992}.
The pressure-dependent reflectivity spectra $R_\text{s-d}$ are affected in the frequency range from 1800 to 2670~cm$^{-1}$ by multi-phonon absorptions in the diamond anvils, which are not completely corrected by the  normalization procedure. This part of the spectrum was interpolated based on the Drude-Lorentz fitting, taking into account the sample–-diamond interface.
The corresponding spectra of the optical conductivity $\sigma(\omega)$=$\sigma_1(\omega)$+i$\sigma_2(\omega)$, dielectric function $\epsilon(\omega)$=$\epsilon_1(\omega)$+i$\epsilon_2(\omega)$, and loss function  -Im(1/$\epsilon$), were obtained by KK analysis of the reflectivty
$R_\text{s-d}(\omega)$ spectra, which were extrapolated to zero based on the Drude-Lorentz fits. For the high-frequency extrapolations, we used the ambient-pressure reflectivity spectrum measured on free-standing crystals, adjusted for the sample-diamond interface. In the KK analysis, the sample–-diamond interface was taken into account as described in Ref.\ \cite{Pashkin2006}.
From the Drude contributions of the Drude-Lorentz fits, the plasma frequency $\omega_{pl}$ was obtained. The plasma frequency and the screened plasma frequency $\omega_{pl}^{scr}$ are related to each other according to $\omega_{pl}^{scr}$=$\omega_{pl}$/$\sqrt{\epsilon_{\infty}}$, where $\epsilon_{\infty}$ is the high-frequency value of $\epsilon_1(\omega)$ accounting for the higher-frequency optical transitions \cite{Schilling.2017}.
At $\omega_{pl}^{scr}$ the loss function has a narrow maximum (plasmon peak) and the function $\epsilon_1(\omega)$ changes its sign \cite{Wooten.1972}.


\begin{figure}[t]
\includegraphics[width=0.4\textwidth]{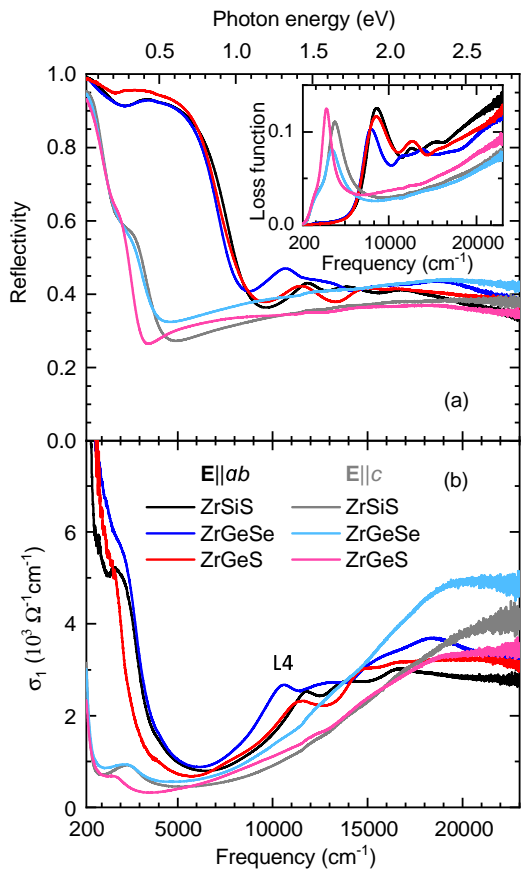}
\caption{(a) Ambient-pressure reflectivity of ZrGeSe, ZrGeS, and ZrSiS along ({\bf E}$\| ab$) and perpendicular ({\bf E}$\| c$) to the square-net layers.
Data of ZrSiS have been reported in Ref.\ \cite{Ebad-Allah.2021}.
Inset: Loss function for the three materials and both polarization directions. (b) Optical conductivity $\sigma_1$ of ZrGeSe, ZrGeS, and ZrSiS for the polarization directions {\bf E}$\| ab$ and {\bf E}$\| c$.}\label{fig:Ambient-pressure}
\end{figure}

\begin{figure}[t]
\includegraphics[width=0.49\textwidth]{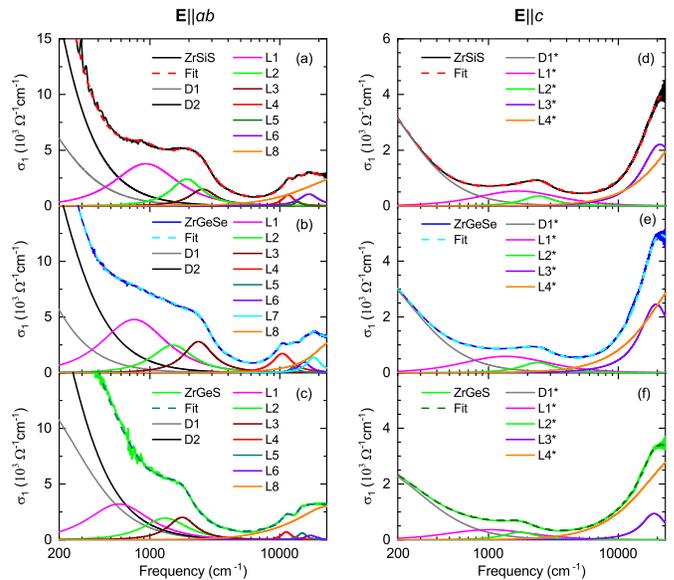}
\caption{Ambient-pressure optical conductivity $\sigma_1$ of ZrSiS, ZrGeSe, and ZrGeS together with the Drude-Lorentz fits and the Drude and Lorentz contributions for the polarization {\bf E}$\| ab$ [(a) - (c)] and {\bf E}$\| c$ [(d) - (f)]. The L4 peak for {\bf E}$\| ab$ is due to transitions between electronic bands related to nonsymmorphic Dirac cones. Results for ZrSiS have been reported in Ref.\ \cite{Ebad-Allah.2021}.}\label{fig:Fitting-contributions}
\end{figure}

\begin{figure}[t]
\includegraphics[width=0.4\textwidth]{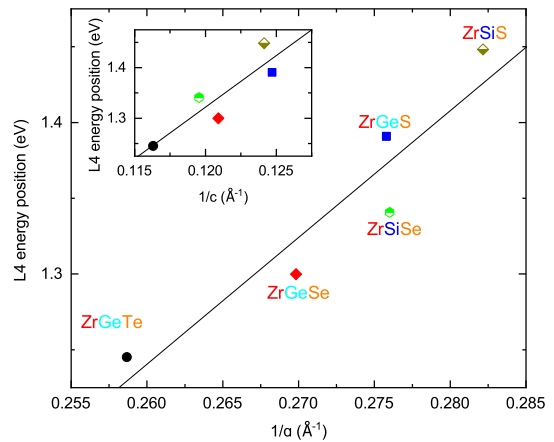}
\caption{L4 energy position as a function of 1/$a$. Inset: L4 energy position as a function of 1/$c$. The values for ZrSiS, ZrSiSe, and ZrGeTe were extracted from Ref.\ \cite{Ebad-Allah.2019}. }\label{fig:L4peakposition}
\end{figure}

\begin{figure*}[t]
\includegraphics[width=1\textwidth]{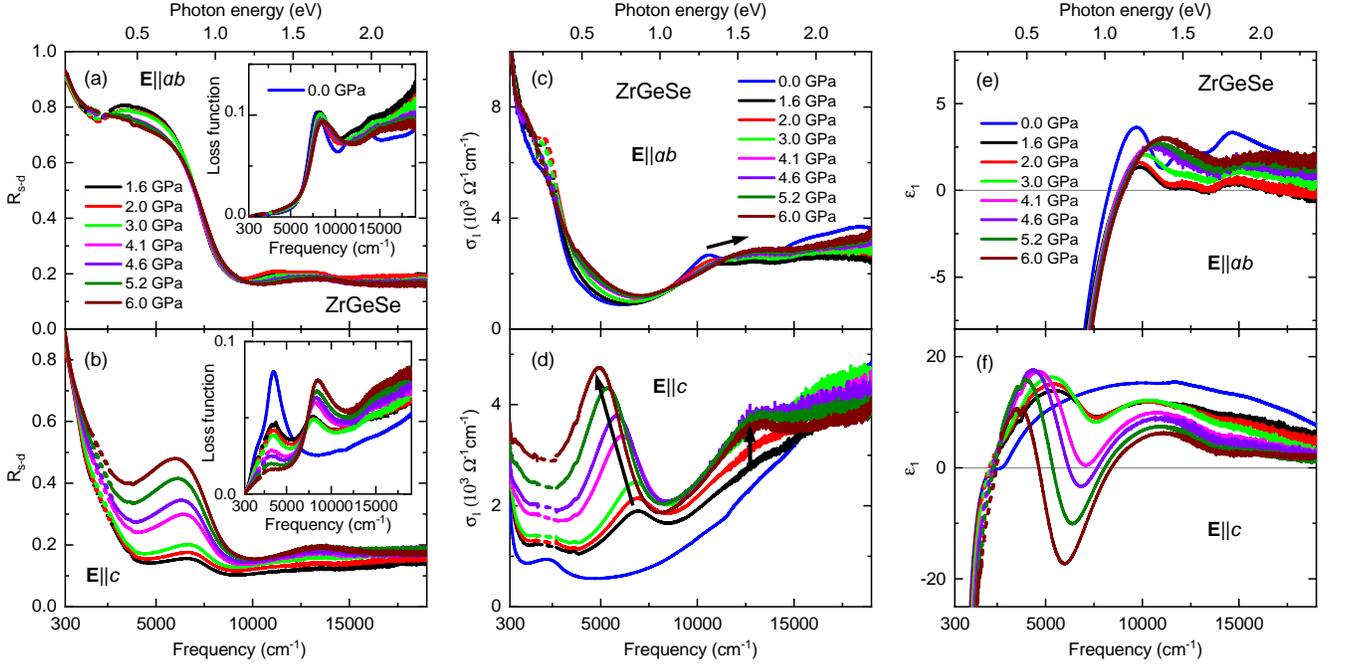}
\caption{(a),(b) Pressure-dependent reflectivity  $R_\text{s-d}$ of ZrGeSe and corresponding pressure-dependent loss function in the inset, for the polarization direction {\bf E}$\| ab$ and {\bf E}$\| c$, respectively. (c),(d) Pressure-dependent optical conductivity $\sigma_1$ of ZrGeSe for the polarization direction {\bf E}$\| ab$ and {\bf E}$\| c$, respectively, as obtained by Kramers-Kronig analysis of the $R_\text{s-d}$ spectra. The arrows highlight the most pronounced pressure-induced changes, in particular the appearance of two new excitations for {\bf E}$\| c$.
(d),(f) Pressure-dependent real part of the dielectric function, $\epsilon_1$, of ZrGeSe for the polarization direction {\bf E}$\| ab$ and {\bf E}$\| c$, respectively.}\label{fig:Pressure-dependent}
\end{figure*}

\begin{figure}[t]
\includegraphics[width=0.4\textwidth]{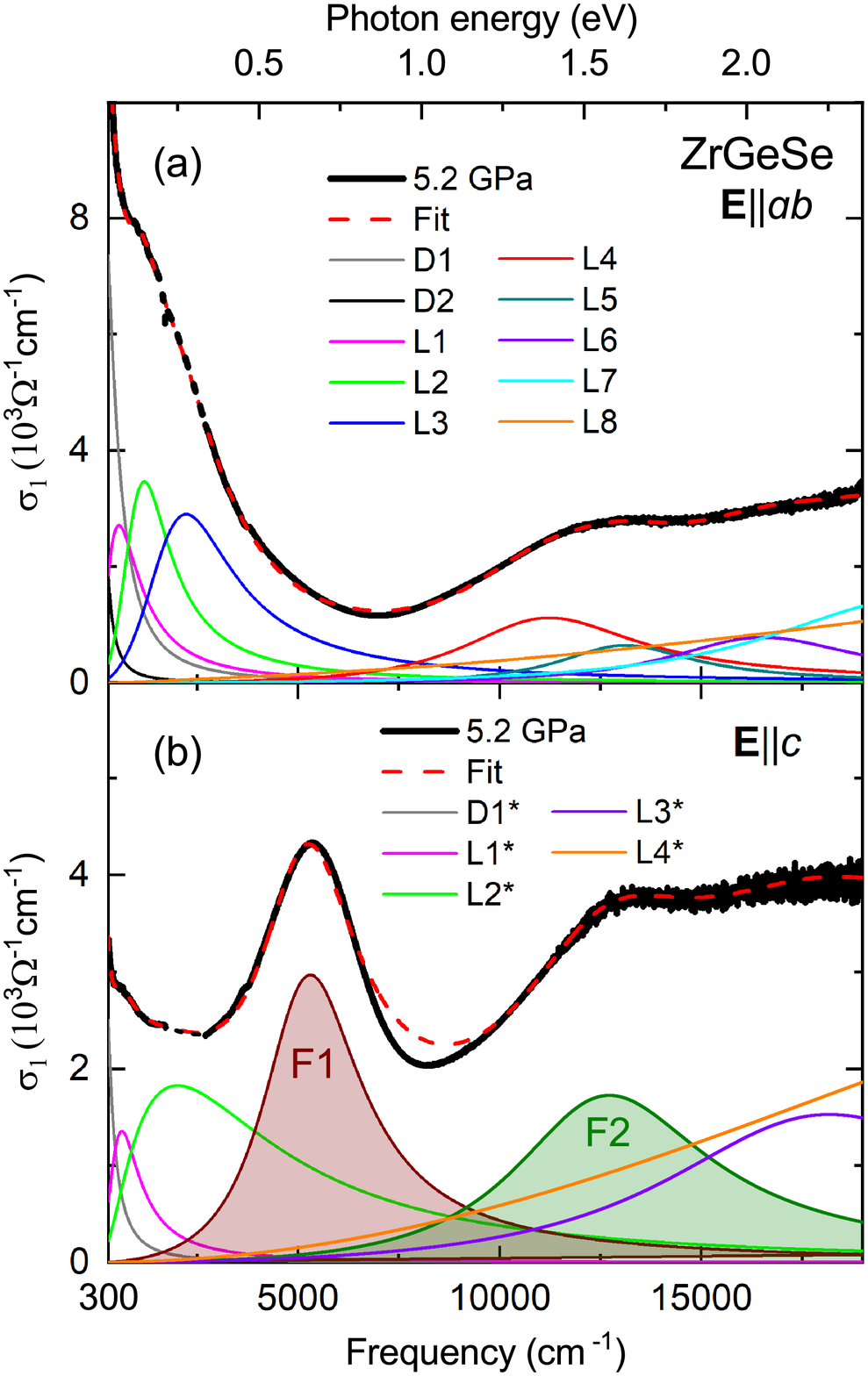}
\caption{Fit of the optical conductivity $\sigma_1$ spectrum of ZrGeSe at 5.2~GPa for (a) {\bf E}$\| ab$ and (b) {\bf E}$\| c$, together with the Drude and Lorentz fitting contributions. F1 and F2 denote the new excitations, which appear under pressure for the polarization {\bf E}$\| c$.}\label{fig:Fithighpressure}
\end{figure}

\section{RESULTS AND DISCUSSION}

\subsection{Ambient-pressure optical response}

The ambient-pressure reflectivity spectra of ZrGeSe, ZrGeS, and ZrSiS are depicted in Fig.\ \ref{fig:Ambient-pressure}(a) for the polarization along ({\bf E}$\| ab$) and perpendicular ({\bf E}$\| c$) to the square-net layers. Results for ZrSiS have already been reported in Ref.\ \cite{Ebad-Allah.2021}.
For all three materials and for both polarization directions, we find a metallic behavior, with a high value of the reflectivity at low frequencies and a distinct plasma edge. The findings for ZrSiS are in agreement with recent electric transport measurements \cite{Novak.2019,Shirer.2019}.
Similarly to the previously observed anisotropic character of ZrSiS \cite{Ebad-Allah.2021}, for both ZrGeSe and ZrGeS the frequency position of the plasma edge is strongly polarization dependent. Namely, for {\bf E}$\| c$ it is shifted towards lower energies compared to {\bf E}$\| ab$. This observation signals the highly anisotropic character of the charge dynamics in all three compounds.
We determined the position of the plasma edge, i.e., the screened plasma frequency $\omega_{pl}^{scr}$, from the corresponding loss function, which shows a plasmon peak at $\omega_{pl}^{scr}$ [see inset of Fig.\ \ref{fig:Ambient-pressure}(a)]. Accordingly, $\omega_{pl}^{scr}$ of ZrSiS, ZrGeSe, ZrGeS is shifted from $\approx$ 1.07, 0.99, and 1.07 eV for {\bf E}$\| ab$ to $\approx$ 0.47, 0.44, and 0.36 eV for {\bf E}$\| c$, respectively (see Table \ref{tab:optical parameters}).


The strongly anisotropic optical response is also revealed by the polarization-dependent optical conductivity $\sigma_1$, as depicted in Fig.\ \ref{fig:Ambient-pressure}(b): Below $\sim$7000~cm$^{-1}$ ($\sim$ 0.87~eV) $\sigma_1$ is much reduced for {\bf E}$\| c$, i.e., perpendicular to the layers.
For {\bf E}$\| ab$, the $\sigma_1$ spectrum consists of (i) Drude contributions followed by (ii) a shoulder or plateau-like behavior up to $\sim$3000~cm$^{-1}$ (0.37~eV) and (iii) a distinct U shape limited by a rather sharp peak (denoted as L4), whose frequency position depends on the material. Please note that the profile of the ambient-pressure in-plane optical conductivity spectrum and its interpretation based on density-functional-theory calculations have been reported in detail in Ref.\  \cite{Ebad-Allah.2019}. Accordingly,
the U-shaped optical conductivity has been ascribed to transitions between linearly crossing bands along a surface in the Brillouin zone, forming an effective nodal plane. The existence of nodal planes is consistent with recent ARPES studies on ZrSiS \cite{Fu.2019}. The shoulder or plateau-like contribution below $\sim$3000~cm$^{-1}$ can be ascribed to transitions within the nodal-line network, which is, in this energy region, affected by corrugation (shift away from the Fermi level) and gapping due to spin-orbit coupling \cite{Ebad-Allah.2019,Kunes.2019}.
The L4 peak is associated with transitions between parallel bands of the Dirac crossings, which are protected by nonsymmorphic symmetry against gapping (these will be denoted as nonsymmorphic Dirac cones in the following) \cite{Ebad-Allah.2019}.
The profile of the {\bf E}$\| c$ optical conductivity spectrum for all three compounds consists of one Drude term followed by a peak-like behavior. The U shape of the profile is, however, not as pronounced as for the in-plane direction, and the L4 peak is missing. In analogy to the in-plane conductivity, we attribute the low-energy optical conductivity to transitions between the linear Dirac bands of the nodal line network.

For ZrGeS, the plateau- and peak-like behavior observed in the {\bf E}$\| ab$ and {\bf E}$\| c$ $\sigma_1$ spectra at low frequencies, respectively, is least developed, i.e., the low-frequency limit of the U-shaped optical conductivity is the lowest among the three compounds. According to Ref.\ \cite{Ebad-Allah.2019}, this signals that ZrGeS is the closest to the ideal nodal-line system.
Indeed, very recent angle-resolved photoemission experiments revealed that the nodal line in ZrGeS shows linear band dispersions within an exceptionally large energy range $>$1.5~eV below $E_F$ \cite{Cheng.2021}.
It is interesting to note that ZrGeS has the lowest $c$/$a$ ratio, i.e., formally this compound has the strongest 3D character and largest interlayer bonding strength \cite{Topp.2016,Ebad-Allah.2019,Ebad-Allah.2019a} among the three materials ZrGeS, ZrGeSe, and ZrSiS.

The optical conductivity and corresponding reflectivity spectra were simultaneously fitted with a phenomenological Drude-Lorentz model (Fig.\ \ref{fig:Fitting-contributions}) for a quantitative analysis and comparison, and for extracting several optical parameters such as the plasma frequency $\omega_{pl}$ (see Table \ref{tab:optical parameters}).
For the polarization {\bf E}$\| ab$, two Drude contributions with $\omega_{pl,1}$ and $\omega_{pl,2}$ had to be inserted, accounting for the electron- and hole-type charge carriers \cite{Lv.2016,Singha.2017}. The value $\omega_{pl}$ was calculated according to $\omega_{pl}$=$\sqrt{\omega^2_{pl,1} + \omega^2_{pl,2}}$ in this case.
The values of the in-plane and out-of-plane $\omega_{pl}$ for ZrSiS are in very good agreement with the values calculated in Ref.\ \cite{Zhou.2019}. Similar values for $\omega_{pl}$ are found for ZrGeS and ZrGeSe.
Furthermore, from the Drude-Lorentz fitting of the {\bf E}$\| ab$  $\sigma_1$ spectrum the energy position of the L4 peak related to the nonsymmorphic Dirac cones was obtained. It has been argued that the relative distance of the nonsymmorphic Dirac cones from $E_F$ is inversely proportional to the distance between two Si atoms in the square net \cite{Kirby.2020}, and hence inversely proportional to the lattice parameter $a$. In order to check this relation, we plot in Fig.\ \ref{fig:L4peakposition} the energy position of the L4 peak, which corresponds to the energy difference between a pair of nonsymmorphic Dirac cones, as a function of 1/$a$. Indeed, we find an approximate linear dependence.
For comparison, the L4 position as a function of the lattice parameter 1/$c$ is given in the inset of  Fig.\ \ref{fig:L4peakposition}. A linear dependence is less obvious in this case, and the reduced $\chi^2$ value (0.0016) of the linear fit is slightly higher than the corresponding value (0.0009) for the 1/$a$ dependence.

For the perpendicular direction {\bf E}$\| c$,
the optical conductivity for frequencies below $\sim$10.000~cm$^{-1}$ is much lower as compared to
{\bf E}$\| ab$ revealing the strongly anisotropic, layered character of all three compounds. The $\sigma_1$ spectrum consists of (i) a Drude term centered at zero frequency followed by (ii) an absorption peak at 1700 - 2400~cm$^{-1}$ (0.2 - 0.3~eV), whose position slightly depends on the material, and (iii) a monotonic increase above 5000~cm$^{-1}$ due to higher-energy interband transitions.
As already discussed above, we ascribe the low-energy optical conductivity to transitions between the linear crossing bands of the nodal line/plane, and the absorption peak to transitions within the corrugated or gapped nodal line network.

\begin{figure*}[t]
\includegraphics[width=0.85\textwidth]{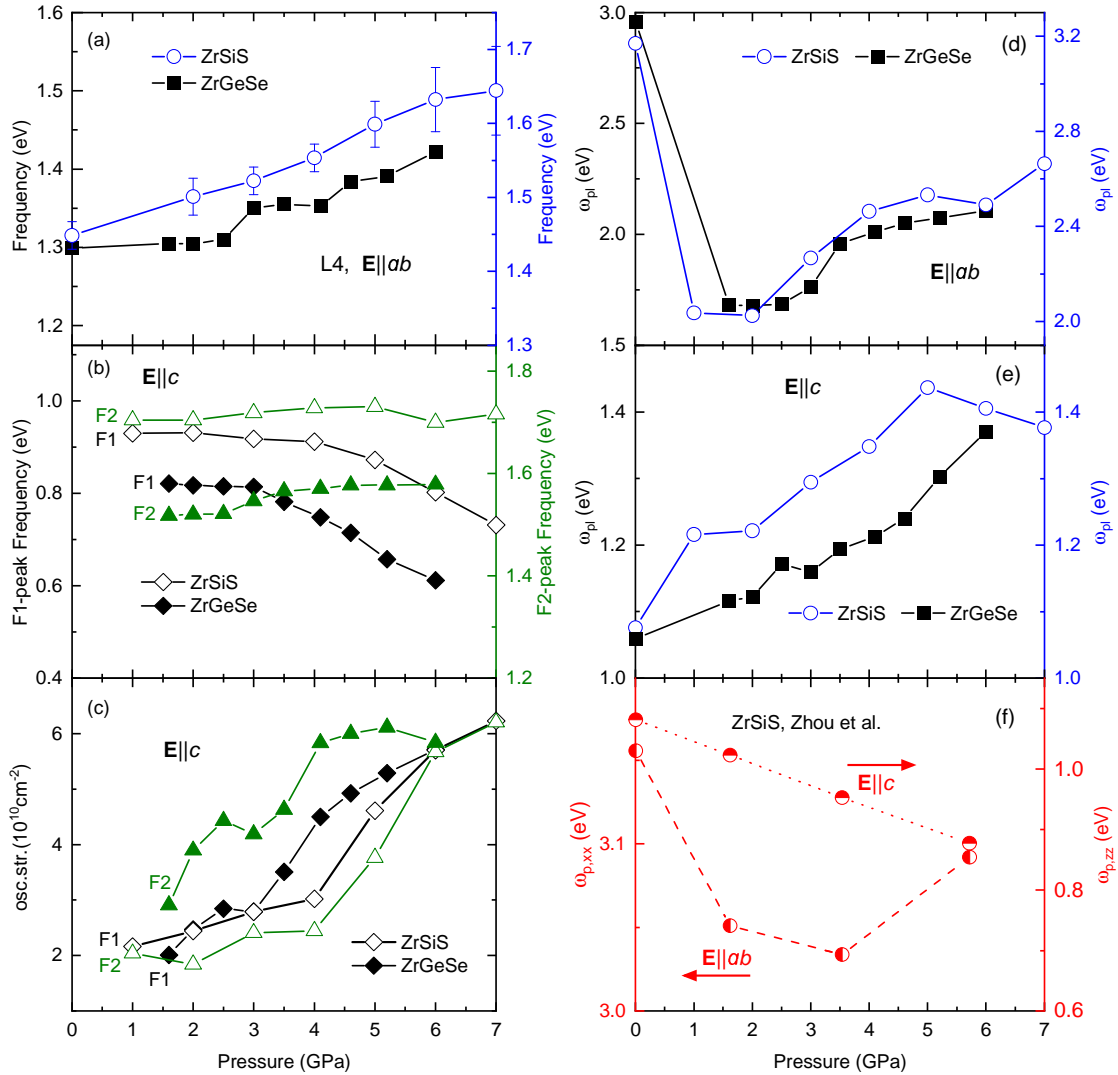}
\caption{(a) Frequency position of the L4 peak for ZrGeSe and ZrSiS as a function of pressure. (b) Frequency position and (c) oscillator strength of the F1 and F2 peak of ZrGeSe and ZrSiS as a function of pressure, respectively. Open (filled) symbols indicate the results for ZrGeSe (ZrSiS).
(d),(e) Plasma frequency of ZrGeSe and ZrSiS as a function of pressure, for {\bf E}$\| ab$ and {\bf E}$\| c$, respectively.
Results for ZrSiS have been extracted from Ref.\ \cite{Ebad-Allah.2021}.
For comparison, the theoretical values of the plasma frequency of ZrSiS under compressive uniaxial strain applied along the $c$ direction from Ref.\ \cite{Zhou.2019} are plotted in (f).
}\label{fig:Fitting-parameters}
\end{figure*}

\subsection{High-pressure optical response}

The previous study on the prototype nodal-line semimetal ZrSiS has revealed pronounced pressure-induced changes in the out-of-plane optical response, with the appearance of two new excitations \cite{Ebad-Allah.2021}. In order to check whether similar pressure-induced changes occur in other compounds of the family, we carried out reflectivity measurements on ZrGeSe as a function of pressure for the polarization directions {\bf E}$\| ab$ and {\bf E}$\| c$ [see Figs.\ \ref{fig:Pressure-dependent}(a) and (b)]. The pressure-induced changes in the {\bf E}$\| ab$ reflectivity are modest, as also revealed by the pressure-dependent loss function displayed in the inset. In contrast, for {\bf E}$\| c$ the profile of the reflectivity spectrum and the loss function change drastically under pressure. These observations are very similar to the recent findings for the polarization-dependent optical response of ZrSiS under pressure \cite{Ebad-Allah.2021}.

The corresponding optical conductivity spectra $\sigma_1$ of ZrGeSe for {\bf E}$\| ab$ and {\bf E}$\| c$ are plotted in Figs.\ \ref{fig:Pressure-dependent}(c) and (d), respectively.
For {\bf E}$\| ab$ the most significant pressure-induced change is the slight shift of the L4 peak to higher energies. For {\bf E}$\| c$ the observed changes are more pronounced, with the appearance of two new excitations, denoted as F1 and F2 in the following.
Additionally, we show in Figs.\ \ref{fig:Pressure-dependent}(e) and (f) the real part of the dielectric function, $\epsilon_1$, of ZrGeSe for both polarization directions at ambient and high pressure. Clearly, the metallic character in both directions is revealed by the large negative values of $\epsilon_1$ at low frequencies and its zero crossing. The pressure-induced changes $\epsilon_1$ are much more drastic for {\bf E}$\| c$ as compared to the in-plane direction. The corresponding results for the polarization-dependent reflectivity, optical conductivity $\sigma_1$ and dielectric function $\epsilon_1$ of ZrSiS under pressure from Ref.\ \cite{Ebad-Allah.2021} are shown in the Supplemental Material \cite{Suppl}.

For a quantitative analysis, the optical conductivity spectra were fitted with a phenomenological Drude-Lorentz model. As examples, we show in Figs.\ \ref{fig:Fithighpressure}(a) and (b) the fits of $\sigma_1$ at 5.2~GPa for both polarization directions, including the Drude and Lorentz contributions.
The main results of this quantitative analysis are summarized in Fig.\ \ref{fig:Fitting-parameters}:
The L4 peak in the {\bf E}$\| ab$ optical conductivity spectrum of ZrGeSe shifts to higher energies with increasing pressure [see Fig.\ \ref{fig:Fitting-parameters}(a)]. This means that the nonsymmorphic Dirac cones shift away from $E_F$ with increasing pressure.
These results are similar to those in ZrSiS \cite{Ebad-Allah.2021}, which we include in Fig.\ \ref{fig:Fitting-parameters}.
The pressure-induced excitations F1 and F2 show a specific frequency shift under pressure [Fig.\ \ref{fig:Fitting-parameters}(b)]. Below $\sim$3~GPa the positions are hardly changed, whereas above $\sim$3~GPa the F1 peak (F2 peak) shifts to lower (higher) frequencies.
The oscillator strength of the peaks as a function of pressure [see Fig.\ \ref{fig:Fitting-parameters}(c)] shows hints for an anomaly at around $\sim$3~GPa, however, it is less obvious than in the frequency shift with pressure.
We also note that the pressure-induced changes in the overall profile of the {\bf E}$\| c$ conductivity spectrum are much enhanced above $\sim$3~GPa [Fig.\ \ref{fig:Pressure-dependent}(d)].
It is interesting to compare these findings with the ones observed for ZrSiS \cite{Ebad-Allah.2021}, which are also included in Fig.\ \ref{fig:Fitting-parameters}(b). For all studied pressures, the F1 and F2 peaks appear at lower frequencies for ZrGeSe as compared to ZrSiS (see Table \ref{tab:optical parameters} for the peak positions at $\sim$1.5 and $\sim$6~GPa). In the case of ZrSiS the anomaly in the parameters of the F1 and F2 peak occurs at $\sim$4~GPa, i.e., at a pressure slightly higher than for ZrGeSe. A higher critical pressure for ZrSiS is consistent with its calculated bulk modulus (144~GPa), which is slightly higher compared to ZrGeSe (121~GPa) \cite{Salmankurt.2016}.
In analogy to ZrSiS, where a pressure-induced structural phase transition could be ruled out \cite{Ebad-Allah.2021}, we suggest that the anomaly at 3~GPa for the F1 and F2 peaks and the conductivity profile observed for ZrGeSe has an electronic origin.

The pressure dependence of the plasma frequency $\omega_{pl}$ for ZrGeSe and ZrSiS, as obtained from the Drude-Lorentz fitting, is displayed in Figs.\  \ref{fig:Fitting-parameters}(d) and (e) for polarization direction {\bf E}$\| ab$ and {\bf E}$\| c$, respectively.
For {\bf E}$\| ab$ one observes an initial drop of $\omega_{pl}$ followed by an increase above $\sim$3~GPa for both compounds.
Qualitatively, this is consistent with the theoretical predictions by Zhou et al.\ \cite{Zhou.2019} based on DFT electronic structure calculations [see Fig.\ \ref{fig:Fitting-parameters}(f)]. Please note that the calculations in Ref.\ \cite{Zhou.2019} were carried out for ZrSiS under compressive uniaxial strain applied along the $c$ direction, which is in principle different from the quasi-hydrostatic conditions in our experiments.
For {\bf E}$\| c$, a linear decrease of $\omega_{pl}$ with increasing compressive strain was predicted for ZrSiS \cite{Zhou.2019}. In contrast, our optical data reveal the opposite behavior for both ZrSiS and ZrGeSe, namely an increase of $\omega_{pl}$ with increasing pressure [Fig.\  \ref{fig:Fitting-parameters}(d)]. Such an increase could be attributed to an enhancement of the interlayer charge carrier transport caused by the pressure-induced decrease in the interlayer distance.

Our pressure-dependent optical data for ZrGeSe thus confirm the recent observation of pressure-induced excitations in the out-of-plane optical response of the closely related compound ZrSiS \cite{Ebad-Allah.2021}. The subtle quantitative differences for the two materials can be explained by the difference in chemical composition, i.e., chemical pressure effect.
The appearance of the pronounced peaks in the {\bf E}$\| c$  optical conductivity spectra of ZrSiS still lacks theoretical explanation.
According to DFT calculations \cite{Ebad-Allah.2021} they cannot be linked to electronic correlation and electron-hole pairing effects in pressurized ZrSiS. It is, however, important to note that, in analogy to the closely related material ZrSiSe \cite{Shao.2020},
electronic correlations are expected to be larger in ZrGeSe as well.
Our confirmation of the pressure-induced excitations in yet another material of the Zr$X$$Y$ compound family will hopefully trigger further theoretical investigations, taking into account additional phenomena such as exciton-polaron formation under pressure, in addition to electronic correlation and electron-hole pairing effects.


\section{CONCLUSION}
The nodal-line semimetals ZrGeSe and ZrGeS, which are closely related to ZrSiS considered as the prototypical nodal-line semimetal, show a pronounced anisotropy in the optical conductivity, with a reduced plasma frequency for the out-of-plane as compared to
the in-plane direction. For both polarization directions, a characteristic profile of the optical conductivity is found, similar to the results for ZrSiS reported in Ref.\ \cite{Ebad-Allah.2021}.
Pressure-dependent experiments on ZrGeSe revealed two new excitations in the out-of-plane optical conductivity
spectrum induced by pressure and anomalies in several optical parameters at $\sim$3~GPa,
confirming the previous findings for ZrSiS.

\begin{acknowledgments}
C.A.K. acknowledges financial support by the Deutsche Forschungsgemeinschaft (DFG), Germany, through grant no.\ KU 1432/13-1. Z.Q.M. acknowledges financial support by the US Department of Energy under grant DE-SC0019068.
\end{acknowledgments}

\end{document}